%% file: Manuscript.tex
\documentclass[final,5p,times]{elsarticle}
\biboptions{numbers,sort&compress}
\usepackage{amsmath,amssymb,bbm,array,amsfonts,float,mathtools,amsthm}
\usepackage{mathrsfs}
\usepackage{booktabs}%
\usepackage{algorithm}
\usepackage{algorithm}%
\usepackage{algorithmicx}%
\usepackage[noend]{algpseudocode}%
\usepackage{listings}%
\usepackage{xcolor}
\usepackage{todonotes}
\usepackage{comment}
\usepackage[hyphens]{url}
\usepackage[hidelinks]{hyperref}
\hypersetup{breaklinks=true}
\usepackage{cjhebrew}
\usepackage[T1]{fontenc}

\input{init}

\journal{Physics Letters B}

\newcommand\ite[1]{\item[\textbf{#1}]}
\newcommand\eref[1]{(\ref{#1})}

\begin{document}
\begin{frontmatter}
\title{Metaheuristic Generation of Brane Tilings\\[1ex]
\small{Dedicated to Professor Amihay Hanany, on the occasion of his 60th birthday}\\
\small\texorpdfstring{\textcjheb{bw.t lzm|}}}

\author[a,b]{Yang-Hui He}
\author[c]{Vishnu Jejjala}
\author[d]{Tom\'as S. R. Silva\corref{cor1}}
\cortext[cor1]{Corresponding author. E-mail address: \href{mailto:tomas@ime.unicamp.br}{tomas@ime.unicamp.br}}

\affiliation[a]{
    organization={London Institute for Mathematical Sciences}, 
    addressline={Royal Institution}, 
    city={London},
    postcode={W1S 4BS},
    country={UK}
}
\affiliation[b]{
    organization={Merton College, University of Oxford}, 
    postcode={OX1 4JD},
    country={UK}
}

\affiliation[c]{
    organization={Mandelstam Institute for Theoretical Physics, School of Physics, NITheCS, \\and CoE-MaSS,
University of the Witwatersrand}, 
    city={Johannesburg},
    postcode={WITS 2050},
    country={South Africa},
}
\affiliation[d]{
    organization={Instituto de Matem\'atica, Estat\'istica e Computa\c{c}\~ao Cient\'ifica (IMECC) da Universidade Estadual de Campinas (UNICAMP)
    }, 
    postcode={13083-859},
    country={Brazil}  
}

\date{\today}

\begin{abstract}
The combinatorics of dimer models on brane tilings describe a large class of four-dimensional $\mathcal{N}=1$ gauge theories that afford quiver descriptions and have toric moduli spaces.
We introduce a combinatorial optimization method leveraging simulated annealing to explicitly construct geometrically consistent brane tilings, providing a proof of concept for efficient generation of gauge theories using metaheuristic techniques. 
The implementation of this idea recovers known examples and allows us to derive a new brane tiling with $26$ quantum fields, illustrating the potential of metaheuristic techniques as a valuable addition to the toolbox for constructing and analyzing gauge theories from brane tilings. 
\end{abstract}

\end{frontmatter}

\input{./Introduction}

\input{./Consistent_tilings}
\input{./Metaheuristic}
\input{./Results}


\section*{Acknowledgements}
We thank Jurgis Pasukonis and Xiao Yan for the code for visualizing the dimers.
YHH is supported by the STFC UK for grant ST/J00037X/2 and a Leverhulme project
grant ``Topology from cosmology: axions, astrophysics and machine learning''.
VJ acknowledges the South African Research Chairs Initiative
of the Department of Science and Innovation and the National Research Foundation.
TSRS is supported by the S\~ao Paulo Research Foundation (FAPESP) grant 2022/09891-4, and highlights that this research utilised computational resources of the ``\textit{Centro Nacional de Processamento de Alto Desempenho em S\~ao Paulo} (CENAPAD-SP)''.

\bibliographystyle{elsarticle-num} 
\bibliography{ref}

\input{./Supplementary_material}

\end{document}

%% file: init.tex
\newcommand{\R}{\mathbb{R}}

\renewcommand{\O}{{\rm O}}

\newcommand{\SU}{{\rm SU}}

\renewcommand{\epsilon}{\varepsilon}

\def\<{\mathopen{}\left<}
\def\>{\right>\mathclose{}}
\def\({\mathopen{}\left(}
\def\){\right)\mathclose{}}

\usepackage{multicol, color}

\definecolor{gold}{rgb}{0.85,.66,0}
\definecolor{cherry}{rgb}{0.9,.1,.2}
\definecolor{burgundy}{rgb}{0.8,.2,.2}
\definecolor{orangered}{rgb}{0.85,.3,0}
\definecolor{orange}{rgb}{0.85,.4,0}
\definecolor{olive}{rgb}{.45,.4,0}
\definecolor{lime}{rgb}{.6,.9,0}
\definecolor{green}{rgb}{.2,.7,0}
\definecolor{grey}{rgb}{.4,.4,.2}
\definecolor{brown}{rgb}{.4,.3,.1}

\newtheorem{theorem}{Theorem}

\theoremstyle{definition}

\newtheorem{example}[theorem]{Example}

%% file: Introduction.tex
\section{Introduction \& summary}
Brane tilings \cite{Hanany:2005ve} reside at the intersection of the physics of supersymmetric quantum field theories \cite{Franco:2005rj,Franco:2005sm}, especially those obtained from string theory, the algebraic geometry of Calabi--Yau manifolds, particularly mirror symmetry and toric geometry \cite{Feng:2000mi,Feng:2005gw,Hanany:2012vc}, and the representation theory of quivers \cite{Beil:2017vku,Franco:2017lpa}.

In its most canonical form, the brane tiling is a geometric and insightful way to encode the data of a $(3+1)$-dimensional supersymmetric gauge theory whose mesonic moduli space is a non-compact toric Calabi--Yau threefold. (See rapid review in \cite{He:2016fnb} and the longer reviews \cite{Kennaway:2007tq, Yamazaki:2008bt}.)
The gauge theory data is a pair $(Q,W)$ where $Q$ is a direct graph (quiver) whose arrows are the fields and whose nodes are the gauge groups; and $W$ is a superpotential which is a polynomial in the fields.
The geometry data is a planar toric diagram --- a convex lattice polygon --- whose associated toric variety is a Fano surface over which the Calabi--Yau threefold is an affine cone.
Realizing these $\mathcal{N}=1$ theories as the worldvolume gauge theory on a D$3$-brane at the tip of a cone over a Sasaki--Einstein base, $X$, we have an infinite class of theories whose gravitational duals are known; they are AdS$_5\times X$ \cite{Klebanov:1998hh, Benvenuti:2004dy}.
The graph dual to $(Q,W)$ is a balanced bipartite graph drawn on $T^2$, known as a brane tiling or dimer model \cite{Franco:2005sm}.
Given the combinatorial nature of the problem, it is natural that there have been catalogues of brane tilings \cite{Davey:2009bp,Hanany:2012hi,Franco:2017jeo,Bao2020}.

The insight of \cite{Jejjala:2010vb,Hanany:2011ra,Hanany:2011bs,He:2012xw,Hanany:2015tgh} was that the data $(Q,W)$ can be succinctly re-packaged into a permutation triple, which weaves Grothendieck's \emph{dessin d'enfants} and Belyi maps into the already rich tapestry.
Here, the physics data is entirely captured by two lists of numbers, bypassing the complexities of graph adjacency or polynomials in arrows of graphs, in the following way.
Fix the permutation group $\mathfrak{S}_d$ on $d$ elements.
Let $\sigma_B$ and $\sigma_W$ be two elements thereof, written explicitly in cycle-notation, and define $\sigma_F = (\sigma_B \sigma_W)^{-1}$.
Then, the gauge theory has $d$ fields with gauge group $\SU(N)^n$ where $n$ is the number of cycles in $\sigma_F$ and the superpotential terms are read off from the cycles in $\sigma_B$ and $\sigma_W$.

The advent of AI driven discovery in theoretical physics and mathematics \cite{He:2017aed,Krefl:2017yox,aggarwal2023, Carifio:2017bov,Ruehle:2017mzq} (see reviews in \cite{He:2018jtw,Ruehle:2020jrk,gukov2024rigor,He:2024gnk}) has naturally led to the consideration of machine learning (ML) applied to the field of quivers, tilings and mutations, in the present context of string theory and in algebraic geometry \cite{He:2020eva,Bao:2020nbi,Bao:2021olg,Arias-Tamargo:2022qgb,Chen:2022jwd,Cheung:2022itk,Dechant:2022ccf,Seong:2023njx,He:2024urx,Armstrong-Williams:2024nzy,Seong:2024wkt,Gukov:2024opc}.
It is therefore expedient to combine our two strands of thought: the efficient permutation triple representation of physics/mathematics, and the possibility of modeling the discovery of new quiver gauge theories as an optimization problem.
This is the chief motivation of our work.
In the remainder of this letter, we describe the encoding of a consistent brane tiling and employ simulated annealing techniques to generate new tilings.

%% file: Consistent_tilings.tex
\section{Consistent tilings}
\label{sec:consistent_tillings}
As discussed in the introduction, brane tilings are efficiently encoded in terms of permutation tuples \cite{Jejjala:2010vb}, whose consistency conditions are presented in~\cite{Hanany:2015tgh}.
In this section we summarize these conditions, with a view towards setting up the data.

Let $d$ be the number of quantum fields.
We take two elements of the symmetric group $\mathfrak{S}_d$, which we call $\sigma_B$ and $\sigma_W$.
The bipartite graph defining the theory has nodes that are colored black and white.
Numbering the edges and reading these anticlockwise at the black nodes and clockwise at the white nodes recovers the two sets of gauge invariant operators that appear in the superpotential.
That is to say, each cycle in $\sigma_B$ and $\sigma_W$ defines one term in the superpotential, and these are assigned couplings $+1$ and $-1$.
As each edge connects a black node to a white node, each field appears twice in the superpotential, once with each sign.
This is the essential requirement for a dimer model.
The defining permutation tuples $(\sigma_B, \sigma_W)$ satisfy the following consistency conditions.
\begin{enumerate}
\ite{PT-1}
Two pairs of tuples $(\sigma_B, \sigma_W)$ and $(\sigma_B', \sigma_W')$ are equivalent when they are related by conjugation:
\begin{equation}
\sigma_B' = \gamma \sigma_B \gamma^{-1} ~, \quad \sigma_W' = \gamma \sigma_W \gamma^{-1} ~,
\end{equation}
for some $\gamma\in \mathfrak{S}_d$.
This ensures that the theory remains the same under a trivial relabeling of the fields.

\ite{PT-2}
The group $G(\sigma_B, \sigma_W)$ whose generators are $\sigma_B$ and $\sigma_W$ is \textit{transitive}.
This means that for any pair of integers $i, j \in\{1,\ldots,d\}$ there exists $g\in G(\sigma_B,\sigma_W)$ such that $g$ sends $i$ to $j$.

\ite{PT-3}
Define
\begin{equation}
\sigma_F := (\sigma_B \sigma_W)^{-1} ~.
\end{equation}
Let $C_\sigma$ count the number of cycles in $\sigma$.
Then, from the Riemann--Hurwitz relations:
\begin{eqnarray}
&& d - C_{\sigma_B} - C_{\sigma_W} - C_{\sigma_F} = 0 ~, \label{eq:rh1} \\
&& C_{\sigma_B} = C_{\sigma_W} ~. \label{eq:rh2}
\end{eqnarray}
Notice that~\eref{eq:rh2} does not imply that the cycle structures of $\sigma_B$ and $\sigma_W$ are the same.

\ite{PT-4}
The lengths of the cycles in $\sigma_B$ and $\sigma_W$ tell us how many fields appear in each term of the superpotential with $\pm$ signs, respectively.
Twice the lengths of the cycles in $\sigma_F$ tell us how many fields bound the faces of the associated bipartite graph.

\ite{PT-5}
There are no one-cycles or two-cycles in either $\sigma_B$ or $\sigma_W$ as these would be tadpoles or mass terms.

\end{enumerate}

We can algorithmically construct the zig-zag paths as an element of $\mathfrak{S}_{2d}$ \cite{Jejjala:2010vb}.
Take the set
\begin{equation}
\Sigma = \{1^+, \ldots, d^+, 1^-, \ldots, d^-\} ~.
\end{equation}
Define a permutation $\mathcal{Z}(\sigma_B,\sigma_W)$ of $\Sigma$ by taking
\begin{equation}
\mathcal{Z}(k^-) = \sigma_B(k)^+ ~, \qquad \mathcal{Z}(k^+) = \sigma_W^{-1}(k)^- ~.
\end{equation}
The cycles in $\mathcal{Z}(\sigma_B,\sigma_W^{-1})$ are the \textit{zig-zag paths}.
The cycles in $\mathcal{Z}(\sigma_B,\sigma_W)$ tell us which fields bound a face in the corresponding bipartite graph.
(As the fields are bifundamental or adjoint, they correspond to edges that separate faces associated to the gauge groups; fixing an orientation, the two faces give a fundamental and an antifundamental index for the field.)

Let $H$ be the Abelian group generated by $\sigma_B\sigma_W^{-1}$.
We require the following two conditions.

\begin{enumerate}

\ite{CONS-1}
For any $h\in H$, $h\sigma_B$ has no one-cycle.
It is called a \textit{derangement}.

\ite{CONS-2}
Choose all pairs $(h_1, h_2)\in H\times H$.
If $h_1 \sigma_B h_2 \sigma_W$ has one-cycles, these belong pairwise to zig-zag paths that are linearly dependent.

\textit{N.B.}: As this is only relevant for generating Seiberg dual phases, it is sufficient to demand that $h_1 \sigma_B h_2 \sigma_W$ be a derangement and then later apply \textit{urban renewal} on the resulting bipartite graph~\cite{Franco:2005rj,Hanany:2011bs}.

\end{enumerate}

For a physical theory, we compute R-charges associated to each of the fields by $a$-maximization~\cite{Intriligator:2003jj}.
These R-charges must satisfy the following conditions.

\begin{enumerate}

\ite{SUPER}
All the terms in the superpotential have R-charge $2$.
Thus, for each cycle in $\sigma_B$ and $\sigma_W$:
\begin{equation}
2 = \sum_{a\in \mathrm{cycle}} R(X_a) ~.
\end{equation}

\ite{NSVZ}
The brane tiling admits an isoradial embedding on a flat torus.
As noted above, the fields in $\mathcal{Z}(\sigma_B, \sigma_W)$ define the faces of the brane tiling.
We require for each face
\begin{equation}
2 = \sum_{a\in \partial F_i} (1-R(X_a)) ~.
\end{equation}
This corresponds to the vanishing of the NSVZ beta functions.

\end{enumerate}

%% file: Metaheuristic.tex
\section{Metaheuristic search for consistent permutation tuples}
We now turn to ML and metaheuristic techniques to see whether we can efficiently generate permutation triples, and hence physical brane tilings, that satisfy all the conditions laid out in the previous section.
After extensive experimentation with the likes of genetic algorithms and reinforcement learning environments, we found that the most efficient method are metaheuristic ones as described in the ensuing.

Metaheuristics are high level probabilistic procedures, often inspired by natural phenomena, designed to find optimal solutions over large search spaces and have proven to be a successful alternative to classical approaches in combinatorial optimization (see~\cite{Bianchi2008} for a review). In particular, \textit{Simulated Annealing} (SA) is a metaheuristic framework introduced in~\cite{Kirkpatrick1983} that relies on a statistical model developed by~\cite{Metropolis1953}. SA emulates the thermodynamic process of \textit{annealing}, representing the dynamics of particles within a heated solid as it progresses toward thermodynamic equilibrium. 

In general, SA is guided by the principle of local search over a space (system) $\mathcal{S}$, using as control parameters a function $\texttt{Energy}:\mathcal{S} \to \R$ associated to any given state in $\mathcal{S}$ and a \textit{temperature} $T^{(i)}$ related to the system in its $i$ -th iteration. The objective is to find a state whose energy approximates the global minimum over the system. To do so, given an initial state in $\mathcal{S}$, at each iteration, a neighbor candidate solution $y$ is computed based on the current solution $x$ according to a function $\texttt{Move}:\mathcal{S} \to \mathcal{S}$. The current solution is updated if $y$ has a lower energy than $x$. However, suppose $y$ has a higher energy than $x$. In that case, the current solution is updated according to a probability distribution that depends on the $i$-th temperature of the system and the energy difference between $x$ and $y$. This probabilistic swap avoids the algorithm being trapped in a local minimum. A pseudocode for a general implementation of SA is presented in Algorithm~\ref{alg:SA} of \ref{appendix:algorithms}.

Consider the problem of searching for geometrically consistent brane tilings encoded as permutation tuples, \textit{i.e.},  finding pairs $(\sigma_B, \sigma_W) \in \mathfrak{S}_d \times \mathfrak{S}_d$ that satisfy the conditions outlined in \S\ref{sec:consistent_tillings}. Among these, \textbf{PT-1} imposes a double-conjugation condition, a standard criterion in permutation group theory to establish equivalence among known consistent tuples. Similarly, \textbf{PT-4} relates the cycle lengths within the tuple to the structure of the induced quiver. However, these conditions do not contribute to the construction of new consistent tuples.

Therefore, to identify geometrically consistent tilings, we proceed as follows. The permutations $\sigma_B$ and $\sigma_W$ are represented as lists of subsets of integers from 1 to $d$, corresponding to their cycle decompositions. The primary goal is to satisfy the conditions \textbf{PT-2}, \textbf{PT-3}, \textbf{PT-5}, \textbf{CONS-1}, and \textbf{CONS-2}. Notably, since any brane tiling that can be mapped onto a torus inherently exhibits locally flat nodes and faces, the \textbf{SUPER} and \textbf{NSVZ} conditions are automatically satisfied~\cite{Hanany:2015tgh}. Our method then searches for permutations that yield a physically valid brane tiling, with \textbf{PT-1} applied at the end to eliminate redundancies.

It is worth noting that a na\"{\i}ve brute force approach to this problem becomes computationally infeasible as the number of quantum fields $d$ increases, with an expected complexity of $\O((d!)^2)$. To address this, we propose a combinatorial optimization method based on simulated annealing (SA) to efficiently search for consistent tilings with large $d$ using finite computational resources.

Therefore, we need to define a suitable energy function that captures how close a given permutation tuple $(\sigma_B, \sigma_W)$ in the search space $\mathcal{S} = \mathfrak{S}_d \times \mathfrak{S}_d$ is from being consistent, so that finding a global minimum for this function corresponds to finding a consistent tiling. This is achieved using routines \texttt{PT-2 Score}, \texttt{PT-3 Score}, \texttt{PT-5 Score}, \texttt{CONS-1 Score}, and \texttt{CONS-2 Score} (detailed in~\ref{appendix:algorithms}), which implement, respectively, scoring routines that measure some \textit{proximity} notion of fulfilling each of \textbf{PT-2}, \textbf{PT-3}, \textbf{PT-5}, \textbf{CONS-1}, and \textbf{CONS-2}; so that a score equals to one is assigned if the correspondent condition is satisfied, and $0 < \texttt{score} < 1$ otherwise.

With these score functions in hand, we can define a function $\texttt{Energy}: \mathfrak{S}_d \times \mathfrak{S}_d \to [0,6)$ as:
\begin{align}
\begin{split}
\label{eq:energy_function}
    \texttt{Energy}(\sigma_B, \sigma_W) = 6 &- (\texttt{PT-2 Score})(\sigma_B, \sigma_W)  \\
    &- (\texttt{PT-3a Score})(\sigma_B, \sigma_W)  \\
    &- (\texttt{PT-3b Score})(\sigma_B, \sigma_W)  \\
    &- (\texttt{PT-5 Score})(\sigma_B, \sigma_W)  \\
    &- (\texttt{CONS-1 Score})(\sigma_B, \sigma_W)  \\
    &- (\texttt{CONS-2 Score})(\sigma_B, \sigma_W) . 
\end{split}
\end{align}

For some $d$, the energy function defined in~\eqref{eq:energy_function} takes a permutation tuple $(\sigma_B, \sigma_W)\in \mathfrak{S}_d \times \mathfrak{S}_d$ and assigns a value in the range $[0, 6)$. The energy associated with a state is zero if, and only if, all consistency conditions are simultaneously satisfied, and it is higher as the number of conditions not (partially) satisfied by a state tuple increases.

It is important to note that during our experiments the partial scoring approach as in~\eqref{eq:energy_function} outperformed binary counting of the total consistency conditions satisfied by a given state. The key difference lies in the smoothness of the energy evaluation between the current solution and a random neighbor. A smoother transition in energy values is crucial for obtaining more reliable estimates of the global minimum of the energy function, corresponding to finding states with zero energy (i.e., consistent tilings).

The remaining component we need to define for executing our SA routine is an appropriate \texttt{Move} function, that produces random neighbors of a given state. There are many possible ways to define such a function in our search space $\mathcal{S}= \mathfrak{S}_d \times \mathfrak{S}_d$ of pairs of permutations, e.g. by randomly shuffling each component of the tuple $(\sigma_B, \sigma_W)$; or by swapping two random elements in each component $\sigma_B$ and $\sigma_W$. However, these do not guarantee the preservation of the cycle structure in the state to be moved and often generate pairs that trivially do not satisfy some cycle-type consistency conditions. Given the $\O((d!)^2)$ nature of our search space, we want to avoid as many trivial unnecessary checks as possible. Thus, we designed a \texttt{Move} algorithm that preserves the cycle structure of a given initial state and therefore does not mishandle previously user-imposed conditions as \textbf{PT-3}~\eqref{eq:rh2} and \textbf{PT-5}. This is exhibited in Algorithm~\ref{alg:move} of \ref{appendix:algorithms}. 

Finally, we mention that both the number of iteration $N$ and the cooling schedule in our SA approach (as in Algorithm~\ref{alg:SA}, \ref{appendix:algorithms}) need to be fine tuned, with optimal performance varying according to the number of quantum fields $d$ that we want to work on. The initial-state $(\sigma_B, \sigma_W)$ cycle structure must also be adjusted. A SageMath~\cite{sagemath} implementation of the routines discussed here is available on the following GitHub repository: \url{https://github.com/TomasSilva/Metaheuristic-brane-tiling-search}.

%% file: Results.tex
\section{Results}
With our SA approach, we were able to reproduce most of the consistent brane tiling presented in previous catalogues~\cite{Davey:2009bp, Franco:2017jeo} up to \textbf{PT-1} equivalence.

In order to generate new examples of tilings, we focused on optimizing our algorithm for larger numbers of quantum fields, for which no complete classification catalogue is known. We emphasize that our metaheuristic SA search for consistent brane tilings with a large number of fields $d$ is still a computationally intensive task since the number of iterations $N$ required by our algorithm to converge to zero-energy states scales significantly with $d$.

Below, we present a new example of consistent brane tiling with $d=26$ quantum fields obtained through our SA approach. This example is not listed (up to equivalence \textbf{PT-1}) in previous catalogues as~\cite{Davey:2009bp, Franco:2017jeo, Bao2020}, which were classifications up to 24 fields.
Thus, this supplies a proof of principle for generating explicit $\mathcal{N}=1$ gauge theories (dimer models) from metaheuristics.

\begin{align}
\begin{split}
\label{eq:26tuples}
 \sigma_B = &(1,3,21,14)(2,4,8)(5,19,24,26)(6,7,25)\\
 &(9,17,11)(10,23,18)(12,22,16)(13,15,20) \\
 \sigma_W = &(1,10,2,6)(3,25,16)(4,12,24)(5,20,23,14)\\
 &(7,13,26)(8,17,15)(9,22,18)(11,19,21)
\end{split}
\end{align}

\begin{figure}[H]
    \centering
    \includegraphics[width=0.75\linewidth]{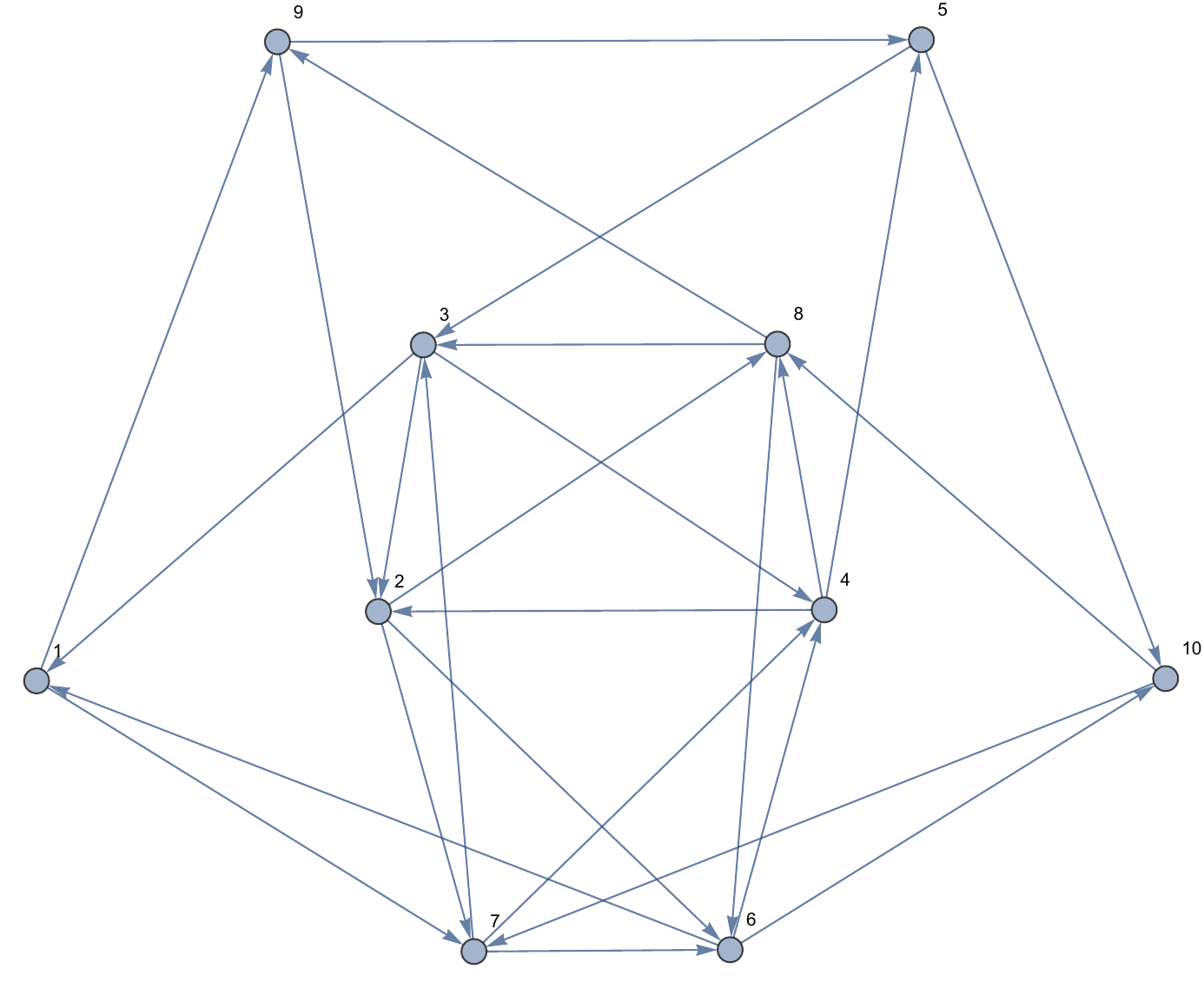}
    \caption{Quiver induced from the permutation tuple in~\eqref{eq:26tuples}.}
    \label{fig:dimer26}
\end{figure}

\begin{figure}[H]
    \centering
    \includegraphics[width=0.7\linewidth]{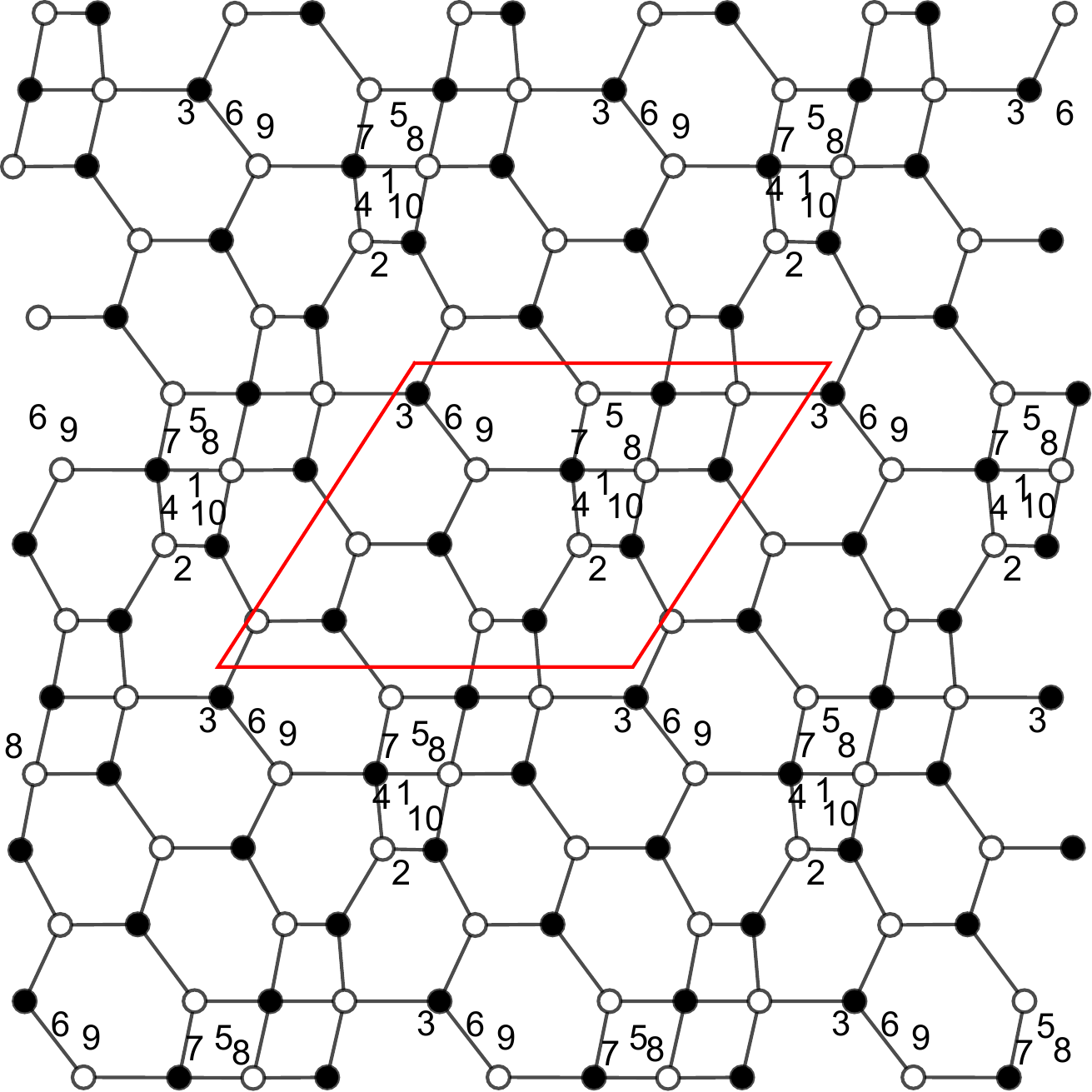}
    \caption{Tiling induced from the permutation tuple in~\eqref{eq:26tuples}.}
    \label{fig:tiling26}
\end{figure}

We can readily write the superpotential from the permutation triples in~\eqref{eq:26tuples}.
First, there are ten gauge group factors, as shown in the quiver diagram in Figure~\ref{fig:dimer26}.
In standard notation $X_{i,j}$ is a bifundamental field from the $i$-th  gauge group $\SU(N_i)$ to the $j$-th (and noting that there are no adjoint fields $\phi_{i,i}$ corresponding to self-joining loops at a node in the quiver), we can write the $16$-term superpontential, consisting of six pairs of cubic interactions and two pairs of quartic, as
\begin{align}
    \begin{split}
        W &= \text{X}_{\text{3,1}}\text{X}_{\text{7,3}}\text{X}_{\text{1,7}}-\text{X}_{\text{2,6}}\text{X}_{\text{4,2}}\text{X}_{\text{6,4}}-\text{X}_{\text{2,7}}\text{X}_{\text{3,2}}\text{X}_{\text{7,3}}+\text{X}_{\text{3,4}}\text{X}_{\text{5,3}}\text{X}_{\text{4,5}}\\
        &-\text{X}_{\text{6,10}}\text{X}_{\text{8,6}}\text{X}_{\text{10,8}}-\text{X}_{\text{6,1}}\text{X}_{\text{7,6}}\text{X}_{\text{1,7}}+\text{X}_{\text{4,2}}\text{X}_{\text{7,4}}\text{X}_{\text{2,7}}+\text{X}_{\text{6,4}}\text{X}_{\text{8,6}}\text{X}_{\text{4,8}}\\        &+\text{X}_{\text{7,6}}\text{X}_{\text{10,7}}\text{X}_{\text{6,10}}+\text{X}_{\text{8,3}}\text{X}_{\text{2,8}}\text{X}_{\text{3,2}}-\text{X}_{\text{9,2}}\text{X}_{\text{8,9}}\text{X}_{\text{2,8}}-\text{X}_{\text{8,3}}\text{X}_{\text{4,8}}\text{X}_{\text{3,4}}\\        &+\text{X}_{\text{1,9}}\text{X}_{\text{6,1}}\text{X}_{\text{2,6}}\text{X}_{\text{9,2}}-\text{X}_{\text{1,9}}\text{X}_{\text{3,1}}\text{X}_{\text{5,3}}\text{X}_{\text{9,5}}\\        &+\text{X}_{\text{5,10}}\text{X}_{\text{9,5}}\text{X}_{\text{8,9}}\text{X}_{\text{10,8}}-\text{X}_{\text{5,10}}\text{X}_{\text{4,5}}\text{X}_{\text{7,4}}\text{X}_{\text{10,7}} \ .
    \end{split}
\end{align}

\begin{figure}[H]
    \centering
    \includegraphics[width=0.6\linewidth]{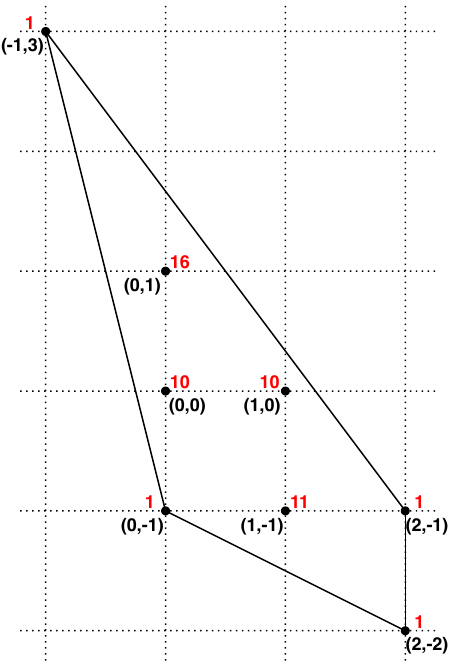}
    \caption{Toric diagram induced from the permutation tuple in~\eqref{eq:26tuples}, with multiplicities in red.}
    \label{fig:toric26}
\end{figure}

The brane tiling, which is the dual graph to the (periodic version of the) quiver together with the superpotential, is drawn as a dimer model on the torus in Figure~\ref{fig:tiling26}, where the red parallelogram marks the fundamental domain.
The moduli space, which can be computed using the forward algorithm, is an affine toric Calabi--Yau threefold, whose planar toric diagram is given in Figure~\ref{fig:toric26}.
We have marked the multiplicity of GLSM fields associated to each point on the toric diagram --- these should count the perfect matchings in the dimer model.
In all, the points of the toric diagram and the corresponding number of perfect matchings are:
\begin{equation}
\begin{array}{c|c}
\mbox{Lattice Point} & \mbox{GLSM/Perfect Matching} \\ \hline
(0, 1) & 16 \\ 
(1, -1) & 11 \\
(0, 0), \ (1, 0) & 10 \\
(2, -2), \ (0, -1), \ (-1, 3), \ (2, -1)
    & 1
\end{array}
\end{equation}
We note that the four external vertices in the toric diagram, \textit{viz.}, $\{(2, -2), \ (0, -1), \ (-1, 3), \ (2, -1)\}$
all have multiplicity 1.
This is consistent with all known theories.
We also note that there are four internal points, and the origin $(0,0)$ is taken to be one of the interior points by convenient choice. Although there are many known brane tilings for toric diagrams with large numbers of internal points (\textit{e.g.}, $Y^{p,q}$~\cite{Franco:2005rj} and $L^{a,b,c}$~\cite{Cveti_2005}), this theory did not appear in any of the previous catalogues~\cite{Davey:2009bp,Hanany:2012hi,Franco:2017jeo,Bao2020} that are primarily concerned with reflexive (single interior point) and two interior points, or with going up in the area of the toric diagram.

We can read off the geometry from the toric diagram and using the toric varieties modules of SageMath~\cite{sagemath}.
Taking the face fan of the lattice polytope, we see that the fan is composed of 4 affine cones, defined cyclically by taking neighbouring rays in $\{
(0,-1), \ (1,-1), \ (2,-1), \ (-1,3)
\}$.
This defines a non-smooth complex surface $B$ overwhich the (Higgs Branch) moduli space is an affine cone which is a non-compact Calabi--Yau threefold $X$.
The (singular) Euler number of $B$ is computed to be $16/5$.

We can construct the cone $X$ as an affine toric variety by adding a coordinate and keep the endpoints co-planar in order to preserve the Calabi--Yau condition.
For instance, we can just append 1 to the two-dimensional lattice points.
This allows one of write down $X$ as a (non-complete) intersection of 72 polynomials in $\mathbb{C}^{12}$.
One can compute the refined Hilbert series for $X$ using~\cite{Benvenuti:2006qr,He:2017gam} directly from reading the toric diagram.
This depends on the triangulation; taking the incidence of the triangulation on the 8 lattice points as $<0,1,3>, <0,3,7>, <1,3,7>, <1,6,7>$, we have
\begin{equation}
    \begin{split}
    HS(X; t_{1,2,3})  &=
    \frac{t_1^2 t_2^2 t_3^4}{\left(1-t_1^2\right) \left(t_1^2 t_2^2-t_3^2\right)
   \left(t_2^2 t_3^2-1\right)}  \\
    &+
       \frac{t_1^4 t_2 t_3^2}{\left(t_1^2-1\right) \left(t_1^2 t_2-t_3\right)
   \left(t_1^4 t_2 t_3-1\right)}  \\
   &-
   \frac{t_1^4 t_2^3 t_3^5}{\left(t_1^2 t_2-t_3\right) \left(t_1^2
   t_2^2-t_3^2\right) \left(t_1^4 t_2^3-t_3^5\right)}\\
   &+
   \frac{t_1 t_2^2 t_3^4}{\left(t_1-t_3^2\right) \left(t_2^2 t_3^2-1\right)
   \left(t_1 t_2^2 t_3^2-1\right)}
   \end{split}
\end{equation}

%% file: Supplementary_material.tex
\appendix
\section{Supplementary material}
\label{appendix:algorithms}
Algorithm~\ref{alg:SA} presents a general implementation of Simulated Annealing.
\begin{algorithm}[H]
    \caption{Simulated Annealing}
    \begin{algorithmic}[1]
        \Require A system $\mathcal{S}$, an energy function $\texttt{Energy}:\mathcal{S} \to \R$, a $\texttt{Move}:\mathcal{S} \to \mathcal{S}$ function, an initial state $x$, an initial temperature $T^{(1)}$, a cooling schedule $\texttt{Cooling}:\R \to \R$, and $N$ the total number of iterations;
        \Ensure An approximate  global minimum of $\texttt{Energy}$ over $\mathcal{S}$;
        \For{$k=1$ \textbf{up to} $N$}
            \State $y = \texttt{Move}(x)$\hfill$\Comment{ \texttt{Move} \text{ provides a neighbour of $x$ in $\mathcal{S}$}}$
            \If{$\texttt{Energy}(y)\leq \texttt{Energy}(x)$}
            \State $x \gets y$
            \ElsIf{$\exp\left( \frac{\texttt{Energy}(x)-\texttt{Energy}(y)}{T^{(k)}}\right)\leq \texttt{URand}([0,1])$}
            \State $x \gets y$
            \EndIf   
            \State $T^{(k+1)} = \texttt{Cooling}(T^{(k)})$$\Comment{\text{Temperature decreases}}$
        \EndFor
    \end{algorithmic}
    \label{alg:SA}
\end{algorithm}

Algorithms \ref{alg:PT2_Score} to \ref{alg:CONS2_Score} provide score evaluations that capture a notion of \textit{being close} to satisfy some consistency condition. They are used to define our SA \texttt{Energy} function, as presented in Eq.~\eqref{eq:energy_function}:

     Algorithm~\ref{alg:PT2_Score} measures how close the group $G(\sigma_B, \sigma_W)$ whose generators are some $\sigma_B$ and $\sigma_W$ is from being transitive (\textbf{PT-2}) by counting its number of orbits. Since a transitive group has only one orbit, this algorithm returns a score $1$ if the group is transitive, and a score in $(0,1)$ if it is not, so that higher number of orbits produces scores closer to $0$.
    \begin{algorithm}[H]
    \caption{\texttt{PT-2 Score}}
    \begin{algorithmic}[1]
       \Require $(\sigma_B, \sigma_W) \in \mathfrak{S}_d \times \mathfrak{S}_d$
       \Ensure 1 if \textbf{PT-2} is satisfied, or a number in $(0,1)$ if it is not.
       \State $G = G(\sigma_B, \sigma_W)$
       \State $n = \# \text{ orbits of } G$
       \State \Return $1/n$
    \end{algorithmic}
    \label{alg:PT2_Score}
\end{algorithm}

    Algorithm~\ref{alg:PT3a_Score} measures how close a tuple $(\sigma_B, \sigma_W)$ is from satisfying the first Riemann--Hurwitz relation (\textbf{PT-3} Eq.~\eqref{eq:rh1}) by computing the value $n:=\left|d - C_{\sigma_B} - C_{\sigma_W} - C_{\sigma_F}\right|$. Since we want $n$ to be zero, a score equal to $1$ is assigned in this case, and a score in $(0,1)$ otherwise, so that higher values of $n$ produce scores closer to $0$.
    \begin{algorithm}[H]
    \caption{\texttt{PT-3a Score}}
    \begin{algorithmic}[1]
       \Require $(\sigma_B, \sigma_W) \in \mathfrak{S}_d \times \mathfrak{S}_d$
       \Ensure 1 if \textbf{PT-3} Eq. \eqref{eq:rh1} is satisfied, or a number in $(0,1)$ if it is not.

       \State $\sigma_F = (\sigma_B \sigma_W)^{-1}$
       \State $C_{\sigma_B}=\#$ cycles of $\sigma_B$
       \State $C_{\sigma_W}=\#$ cycles of $\sigma_W$
       \State $C_{\sigma_F}=\#$ cycles of $\sigma_F$
       
       \State $n=\left|d-C_{\sigma_B}-C_{\sigma_W}-C_{\sigma_F}\right|$

       \State \Return $1/(n+1)$
    \end{algorithmic}
    \label{alg:PT3a_Score}
\end{algorithm}

     Algorithm~\ref{alg:PT3b_Score} measures how close a tuple $(\sigma_B, \sigma_W)$ is from satisfying the second Riemann--Hurwitz relation (\textbf{PT-3} Eq.~\eqref{eq:rh2}) by computing the value $n:=\left|C_{\sigma_B} - C_{\sigma_W}\right|$. Since we want $n$ to be zero, a score equal to $1$ is assigned in this case, and a score in $(0,1)$ is assigned otherwise, so that higher values of $n$ produce scores closer to $0$.
    \begin{algorithm}[H]
    \caption{\texttt{PT-3b Score}}
    \begin{algorithmic}[1]
       \Require $(\sigma_B, \sigma_W) \in \mathfrak{S}_d \times \mathfrak{S}_d$
       \Ensure 1 if \textbf{PT-3} Eq. \eqref{eq:rh2} is satisfied, or a number in $(0,1)$ if it is not.

       \State $C_{\sigma_B}=\#$ cycles of $\sigma_B$
       \State $C_{\sigma_W}=\#$ cycles of $\sigma_W$

       \State $n=\left|C_{\sigma_B}-C_{\sigma_W}\right|$

       \State \Return $1/(n+1)$
    \end{algorithmic}
    \label{alg:PT3b_Score}
\end{algorithm}

     Algorithm~\ref{alg:PT5_Score} measures how close a tuple $(\sigma_B, \sigma_W)$ is from not having one-cycles or two-cycles in either $\sigma_B$ or $\sigma_W$ (\textbf{PT-5}) by computing the value $n:=C^{(1)}_{\sigma_B}+C^{(1)}_{\sigma_W}+C^{(2)}_{\sigma_B}+C^{(2)}_{\sigma_W}$, the sum of one-cycles and two-cycles in both $\sigma_B$ and $\sigma_W$. Since we want $n$ to be zero, a score equal to $1$ is assigned in this case, and a score in $(0,1)$ is assigned otherwise, so that higher values of $n$ produce scores closer to $0$.
    \begin{algorithm}[H]
    \caption{\texttt{PT-5 Score}}
    \begin{algorithmic}[1]
       \Require $(\sigma_B, \sigma_W) \in \mathfrak{S}_d \times \mathfrak{S}_d$
       \Ensure 1 if \textbf{PT-5} is satisfied, or a number in $(0,1)$ if it is not.
       \State $C^{(1)}_{\sigma_B}=\#$ one-cycles in $\sigma_B$
       \State $C^{(1)}_{\sigma_W}=\#$ one-cycles in $\sigma_W$
       \State $C^{(2)}_{\sigma_B}=\#$ two-cycles in $\sigma_B$
       \State $C^{(2)}_{\sigma_W}=\#$ two-cycles in $\sigma_W$
       \State $n=C^{(1)}_{\sigma_B}+C^{(1)}_{\sigma_W}+C^{(2)}_{\sigma_B}+C^{(2)}_{\sigma_W}$
       \State \Return $1/(n+1)$
    \end{algorithmic}
    \label{alg:PT5_Score}
\end{algorithm}

     Algorithm~\ref{alg:CONS1_Score} measures how close all possible $h\sigma_B$ are from not having one-cycles, where $h$ runs all over the Abelian group generated by $\sigma_B\sigma_W^{-1}$ (\textbf{CONS-1}). This is done by summing up the numbers of one-cycle obtained in all possible $h\sigma_B$. Since we want this summation to be zero, a score equal to $1$ is assigned in this case, and a score in $(0,1)$ is assigned otherwise, so that a higher number of one-cycles detected in all possible pairings produces scores closer to $0$;
    \begin{algorithm}[H]
    \caption{\texttt{CONS-1 Score}}
    \begin{algorithmic}[1]
       \Require $(\sigma_B, \sigma_W) \in \mathfrak{S}_d \times \mathfrak{S}_d$
       \Ensure 1 if \textbf{CONS-1} is satisfied, or a number in $(0,1)$ if it is not.
       \State $H = \langle \sigma_B\sigma_W^{-1}\rangle$
       \State $n=0$
       \For{$h \in H$}
        \State $\texttt{n1c}=\#$ one cycles in $h\sigma_B$
        \State $n\gets n+\texttt{n1c}$
       \EndFor
       \State \Return $1/(n+1)$
    \end{algorithmic}
    \label{alg:CONS1_Score}
\end{algorithm}

     Algorithm~\ref{alg:CONS2_Score} measures how close all possible $h_1\sigma_B h_2 \sigma_W$ are from not having one-cycles, where $h_1$ and $h_2$ runs all over the Abelian group generated by $\sigma_B\sigma_W^{-1}$ (\textbf{CONS-2}). This is done by summing up the numbers of one-cycles obtained in all possible $h_1\sigma_B h_2 \sigma_W$. Since we want this summation to be zero, a score equal to $1$ is assigned in this case, and a score in $(0,1)$ is assigned otherwise, so that a higher number of one-cycles detected in all possible pairings produces scores closer to $0$.
\begin{algorithm}[H]
    \caption{\texttt{CONS-2 Score}}
    \begin{algorithmic}[1]
       \Require $(\sigma_B, \sigma_W) \in \mathfrak{S}_d \times \mathfrak{S}_d$
       \Ensure 1 if \textbf{CONS-2} is satisfied, or a number in $(0,1)$ if it is not.
       \State $H = \langle \sigma_B\sigma_W^{-1}\rangle$
       \State $n=0$
       \For{$(h_1, h_2) \in H \times H$}
        \State $\texttt{n1c}=\#$ one cycles in $h_1\sigma_B h_2 \sigma_W$
        \State $n\gets n+\texttt{n1c}$
       \EndFor
       \State \Return $1/(n+1)$
    \end{algorithmic}
    \label{alg:CONS2_Score}
\end{algorithm}

Algorithm~\ref{alg:move} details our cycle-type preserving \texttt{Move} function. We also illustrate its execution in Example~\ref{ex:move_alg}.

\begin{algorithm}[H]
    \caption{\texttt{Move}}
    \begin{algorithmic}[1]
       \Require $(\sigma_B, \sigma_W) \in \mathfrak{S}_d \times \mathfrak{S}_d$
       \Ensure $(\widetilde{\sigma_B}, \widetilde{\sigma_W}) \in \mathfrak{S}_d \times \mathfrak{S}_d$ with same cycle type of $(\sigma_B, \sigma_W)$
       \State $c_{i_B} , c_{j_B}$ = two random cycles in $\sigma_B$ 
       \State $c_{i_W} , c_{j_W}$ = two random cycles in $\sigma_W$ 

       \State $i_B$ = random element in cycle $c_{i_B}$ of $\sigma_B$
       \State $j_B$ = random element in cycle $c_{j_B}$ of $\sigma_B$
       \State $i_W$ = random element in cycle $c_{i_W}$ of $\sigma_W$
       \State $j_W$ = random element in cycle $c_{j_W}$ of $\sigma_W$

        \State $\widetilde{\sigma_B}$ = swap element $i_B$ in cycle $c_{i_B}$ with element $j_B$ in cycle $c_{j_B}$ of $\sigma_B$
        
        \State $\widetilde{\sigma_W}$ = swap element $i_W$ in cycle $c_{i_W}$ with element $j_W$ in cycle $c_{j_W}$ of $\sigma_W$

        \State \Return $(\widetilde{\sigma_B}, \widetilde{\sigma_W})$
    \end{algorithmic}
    \label{alg:move}
\end{algorithm}

\begin{example}
\label{ex:move_alg}
Below we exemplify the execution of  Algorithm~\ref{alg:move}, \texttt{Move}, applied to a given permutation tuple in $\mathfrak{S}_{8}\times \mathfrak{S}_{8}$:
\begin{enumerate}
    \item Consider the following two initial permutations presented in cycle notation:
    \[
    \begin{split}
    \sigma_B = (1,3,4,7)(2,8,6,5) \\
    \sigma_W = (1,8,2)(3,6,5,4,7)
    \end{split}
    \]
    The cycle type of $\sigma_B$ is $(4,4)$ and the cycle type of $\sigma_W$ is $(3,5)$. Note that $C_{\sigma_B} = C_{\sigma_W}=2$, and that there are no one-cycles or two-cycles, so that \textbf{PT-3} Eq.~\eqref{eq:rh2} and \textbf{PT-5} are satisfied by construction.

    \item Choose two random cycles in each $\sigma_B$ and $\sigma_W$:
    \[
    \begin{split}
        \sigma_B = \underbrace{(1,3,4,7)}_{c_{i_B}}\underbrace{(2,8,6,5)}_{c_{j_B}}\\
        \sigma_W = (1,8,2)\underbrace{(3,6,5,4,7)}_{c_{i_W}=c_{j_W}}
    \end{split}
    \]
    \item Choose a random element in each of the chosen cycles:
    \[
    \begin{split}
    \sigma_B = (1,\textcolor{red}{\underset{i_B}{\underset{\uparrow}{3}}},4,7)(2,8,6,\textcolor{blue}{\underset{j_B}{\underset{\uparrow}{5}}})\\
    \sigma_W = (1,8,2)(1, \textcolor{red}{\underset{i_W}{\underset{\uparrow}{6}}}, 5, \textcolor{blue}{\underset{j_W}{\underset{\uparrow}{4}}},7)
    \end{split}
    \]

    \item Do the swap:
    \[
    \begin{split}
    \widetilde{\sigma_B} = (1,\textcolor{blue}{5},4,7)(2,8,6,\textcolor{red}{3}) \\
    \widetilde{\sigma_W} = (1,8,2)(3,\textcolor{blue}{4},5,\textcolor{red}{6},7)
    \end{split}
    \]
    Note that after the \texttt{Move} execution, both $\widetilde{\sigma_B}$ and $\widetilde{\sigma_W}$ preserve the initial cycle type of $\sigma_B$ and $\sigma_W$, respectively, $(4,4)$ and $(3,5)$. 
\end{enumerate}
\end{example}